\documentclass{article}
\usepackage{arxiv}
\usepackage{algorithm}
\usepackage{wrapfig}
\usepackage[dvipsnames]{xcolor}
\usepackage[utf8]{inputenc} 
\usepackage[T1]{fontenc}    
\usepackage{hyperref}       
\usepackage{url}            
\usepackage{booktabs}       
\usepackage{amsfonts}       
\usepackage{nicefrac}       
\usepackage{microtype}      
\usepackage{lipsum}
\usepackage{longtable}
\usepackage{multicol}
\usepackage{graphicx,algorithm,algorithmic}
\usepackage{float}
\usepackage{multirow}
\usepackage[caption = false]{subfig}
\usepackage{adjustbox}

\newcommand{\sitevec}{{\textbf{Site2Vec}}}
\title{Site2Vec: a reference frame invariant algorithm for vector embedding of protein-ligand binding sites}

\author{
  Arnab Bhadra \\
  Computer Science and Engineering\\
  IIT Tirupati\\
  India, 517506\\
  \texttt{cs18s501@iittp.ac.in} \\
   \And
 Kalidas Yeturu \\
  Computer Science and Engineering\\
  IIT Tirupati\\
  India, 517506\\
  \texttt{ykalidas@iittp.ac.in} \\
}
\date{\vspace{-5ex}}
\begin{document}

\maketitle

\vspace{10pt}

\begin{abstract}
Protein-ligand interactions are one of the fundamental types of molecular interactions in living systems. 
Ligands are small molecules that interact with protein molecules at specific regions on their surfaces called binding sites.
Binding sites would also determine ADMET properties of a drug molecule.
Tasks such as assessment of protein functional similarity and detection of side effects of drugs need identification of similar binding sites of disparate proteins across diverse pathways. 
To this end, methods for computing similarities between binding sites are still evolving and is an active area of research even today.
Machine learning methods for similarity assessment require feature descriptors of binding sites.
Traditional methods based on hand engineered motifs and atomic configurations are not scalable across several thousands of sites.
In this regard, deep neural network algorithms are now deployed which can capture very complex input feature space.
However, one fundamental challenge in applying deep learning to structures of binding sites is the input representation and the reference frame.
We report here a novel algorithm \sitevec\ that derives reference frame invariant vector embedding of a protein-ligand binding site.
The method is based on pairwise distances between representative points and chemical compositions in terms of constituent amino acids of a site.
The vector embedding serves as a locality sensitive hash function for proximity queries and determining similar sites.
The method has been the top performer with more than 95\% quality scores in extensive benchmarking studies carried over 10 data sets and against 23 other site comparison methods of the field.
The algorithm serves for high throughput processing and has been evaluated for stability with respect to reference frame shifts, coordinate perturbations and residue mutations.
We also provision the method as a stand alone executable and a web service hosted at \url{http://services.iittp.ac.in/bioinfo/home}.

\end{abstract}

%
\vspace{2pc}

\section{Introduction}
\label{introduction}

A protein performs its function by interacting with ligands and other small molecules. 
Protein-ligand interaction plays an important role in a biological system.
A binding site is a concavity on the surface of a protein, composed of several amino acids whose side chains interact with a ligand molecule.
Electrostatic and chemical complimentarity, aromatic stacking and other forms of interactions between atoms of a concavity and a ligand molecule result in chemical binding.
One protein may have a dozen concavities and some of which may be functional resulting in moonlighting behaviour.
A major reason for functional similarity between proteins is the similarity of their binding sites and hence interaction patterns with ligand molecules.
Binding site comparison is one of the major methods in the field of structural bioinformatics and drug discovery \cite{xie2009drug, goh2017deep}.

A main source for structure of a binding site, is the structure of its protein in terms of 3D coordinates of all of its thousands of atoms.
The structure of a protein is determined through a variety of means including experimental mechanisms such as X-ray crystallography or nuclear magnetic resonance and analytical methods such as sequence or structural alignment to known structures \cite{Kuhlman2019}.
Today about 167 thousand high quality protein structures are available in the Protein Data Bank \cite{berman2003protein}.
This volume of data has enabled use of machine learning methodology for comparison of protein-ligand binding sites \cite{krivak2018p2rank, hassan2018dlscore}. 
Traditional machine learning methodologies over the last two decades require careful feature engineering and sometimes involve hand crafted features and atomic motifs.
However for scaling to handle thousands of features originating in data such as medical text, protein function, pathway information and other sources, deep learning methodologies have come into picture \cite{chen2018rise, mayr2016deeptox, ragoza2017protein}.

There are two types of site comparison methods, alignment-dependent methods \cite{gao2013apoc, yeturu2011pocketalign,konc2010probis,chartier2015detection,zhang2005tm} and alignment-free methods \cite{weill2010alignment,nakamura2016protein,krotzky2015large,wood2012pharmacophore,yeturu2008pocketmatch}.
Alignment dependent methods provide detailed information on the atomic mapping of protein structures where as alignment free methods output a similarity score.
Several methods have come till date for the purpose of binding site comparison \cite{weill2010alignment,chartier2015detection,desaphy2013encoding,wood2012pharmacophore,yeturu2008pocketmatch,konc2010probis,krotzky2015large,zhang2005tm,batista2014sitehopper,shulman2005siteengines,schalon2008simple,gao2013apoc,xie2008detecting,brylinski2014matchsite,caprari2014assist,desaphy2012comparison,morris2005real,jimenez2017deepsite,nakamura2016protein,yeturu2011pocketalign,morris2005real,pu2019deepdrug3d,gold2006sitesbase}.
A brief description of these methods is provided in (Table~\ref{tab:literature}).

\begin{longtable}{c|p{9cm}}
    \caption{Brief descriptions of methods for binding site comparison} \\ \hline
             \textbf{Method} &  \textbf{Description} \\ \hline
             Apoc \cite{gao2013apoc} & Binding sites are aligned based on matching residue groups between sites and carrying out site alignment. \\
             ASSIST \cite{caprari2014assist} & It is a geometric hashing based comparison technique, hashes pairs of residues, taken from input structures into a 3D hash table based on residue types and distances from geometric centers. Similarity exercise is performed through alignment of hash tables.\\
             DeepDrug3D \cite{pu2019deepdrug3d} & A binding site is encapsulated in a 3D voxel-grid which is passed through a convolutional neural network to map to a vector descriptor. The method falls under supervised learning and requires categories of sites as input.\\
             eMatchSite \cite{brylinski2014matchsite} & A sequence-order independent algorithm for binding site comparison. Physico-chemical and structural characteristics of protein residues and their interactions with small molecules are captured by the spatial distribution of residues and ligand binding probabilities. \\   
             FuzCav \cite{weill2010alignment} &  An alignment-free comparison method that represents a binding site as a 4833 integer vector of counts of the number of pharmacophoric triplets formed by $C_\alpha$ atoms.\\ 
             IsoMIF \cite{chartier2015detection} & A Molecular interaction-based binding site comparison method where a binding site is represented by a molecular interaction field descriptor. Interactions between atoms of residues of a site and a ligand in a protein ligand complex are input to the method. \\  
             KRIPO \cite{wood2012pharmacophore} & Quantifying the similarities at the level of sub-pockets in a binding site based on pharmacophore representation. The presence or absence of sub-pockets is then then translated to derive a fingerprint. \\ 
			 Morris and co-workers \cite{morris2005real} & A binding site is represented by spherical harmonic coefficients about its centroid.\\       
             Nakumara and co-workers \cite{nakamura2016protein} & In their method, a binding site is represented by a 11-dimensional vector, reducing the fingerprint obtained by FuzCav \cite{weill2010alignment} using principal component analysis on 4833 dimensional space.  \\
             PocketAlign \cite{yeturu2011pocketalign} & A pair of binding sites are structurally superimposed based on matching residues for their local shape descriptors between sites. \\             
             PocketMatch \cite{yeturu2008pocketmatch} & Represents the binding site as lists containing distances between $C_{\alpha}$, $C_{\beta}$ and centroid of atoms of side chain and categorized into physico-chemical property bins. A pair of sites are compared for similarity by matching distance elements between corresponding bins.\\
             Probis \cite{konc2010probis} & In this method a site is represented as a fully connected graph whose vertices are atoms. Two sites are compared based on maximal common sub-graph detection algorithms. \\
             RAPMAD \cite{krotzky2015large} & A histogram-based comparison method where several sets of pseudo centers represent a binding site.  Distances between pseudo centers are calculated and mapped into histograms. These histograms represent binding sites which are compared for similarity. \\
             SiteAlign \cite{schalon2008simple} & A binding site is represented by a sphere containing physico-chemical properties on its surface. Site residues are projected on to this sphere. Two sites are compared by alignment of their sphere approximations. \\
             SiteHopper \cite{batista2014sitehopper} & It calculates a 3D shape representation of the active site colored by the chemical properties of the  residues defining the active site. These active site representations are aligned and assessed for shape and  chemical similarity. \\
             SiteEngine \cite{shulman2005siteengines} & Each amino acid in the site is represented by a pseudo-centre with physico-chemical label. two sites are superimposed for matching pseudo centers.\\  
             SitesBase \cite{gold2006sitesbase} & Alignment based algorithm that superimposes atoms between a pair of sites. Alignments are pre-computed and served through a database.\\                                                            	
             SOIPPA \cite{xie2008detecting} & Represents protein structures by Delaunay tessellation of $C_\alpha$ atoms characterized by their geometric potentials. Two structures are aligned by determining maximal weight common sub-graph. \\
             TM-align \cite{zhang2005tm} & An Alignment dependent approach where a protein structure is treated as position of $C_{\alpha}$ atoms of residues. Two structures are aligned by superimposing their backbone atoms. \\
             VolSite \cite{desaphy2012comparison} & A site is approximated by fitting a box into the pocket and lining the surface of the box with pharmacophoric properties. Two sites are compared through alignment of these box representations. \\            

    \label{tab:literature}
\end{longtable}

Pairwise alignment based methods need a {\bf pair of sites} as input.
Representative points in 3D are mapped between the sites and a superimposition is carried out.
PocketAlign \cite{yeturu2011pocketalign} is an example of alignment based approach.
Results of alignment provided detailed information on atomic mapping and can be visualized in standard applications.
These algorithms also emit a matching score on number of atoms or residues matched between the sites.

While alignment is a detailed work for a given pair of sites, dealing with several million pairs is space and time complexity wise hard.
A precomputed database of alignments and scores are stored to serve for a long list of known sites \cite{gold2006sitesbase}.
However this approach does not scale well for querying a large number of novel sites.
Especially in protein design experiments for synthetic enzymes and pharmacophore based generation of binding sites \cite{syntheticEnzyme2019}.
Moreover in the context of pipelines of drug discovery processes involving screening against thousands of sites for side effect free ligand selection \cite{Raman2008TargetTB}, more than a detailed alignment based approach, a quick and accurate filter is desirable.

The advantage of a vector descriptor based algorithm is its ability for use in diverse contexts.
Queries for identification of similar binding sites reduces to proximity search in their vector descriptor space.
Classifying a pair of binding sites into a category is achieved through concatenation of vectors and processing the fused vector through machine learning classifiers. 

In the space of {\it pairwise methods}, PocketMatch \cite{yeturu2008pocketmatch} has been fastest \cite{ehrt2018benchmark} and competitively accurate across diverse data sets.
PocketMatch considers a binding site structure as composed of amino acids represented by 3 point-types $C_\alpha, C_\beta$ and $C_\gamma$ (centroid of side chain atoms).
The twenty amino acids are grouped into 5 chemical categories.
All pair distances between the points are then mapped into 120 bins corresponding to point-chemical-type pairs and sorted within each bin.
This is the shape descriptor or finger print for a binding site.

However, PocketMatch bins are variable in length being proportional to the number of atoms and characteristic chemical composition of a site.
Though the representation serves the purpose in pairwise scanario, the variable vector is not suitable for proximity queries.
In this work \sitevec, we mathematically extend the idea of PocketMatch to generate a fixed size vector which can serve as a locality-sensitive-hash descriptor.

In \sitevec, a sliding window of numbers runs through each physico-chemical-bin of PocketMatch.
The slides are all clustered for each physico-chemical-bin to form anchor points.
For any new site, sliding window is run for all the bins and a histogram of cluster membership is derived.
These histograms are then encoded using auto-encoder to form a $200$-dimensional vector descriptor.

The algorithm has been benchmarked against 23 methods on 10 data sets.
On average \sitevec\ stood first in its performance for classification of sites provided in recent extensive benchmarking paper \cite{ehrt2018benchmark}.
The algorithm is also robust to atomic coordinate perturbations characteristic of any site, single amino acid mutations and reference frame changes.
The algorithm is fast which computes a vector descriptor at the rate of 3-5 sites per second, making it suitable for high throughput processing.

There has been recent work that employs deep learning for vectorization of binding sites \cite{pu2019deepdrug3d}.
In this method a voxel representation of 3D space featuring interaction energy of 14 atom types of a binding site is passed through a convolutional neural network (CNN) to obtain a phyisico-chemical shape descriptor.
As CNNs are sensitive to reference frame, in this algorithm, a pre-processing step is carried out to determine the three axes using principal component analysis.
However one major drawback of the algorithm is, as CNN is a supervised learning algorithm, it requires site categories beforehand and is not suitable in a large scale scenario across hundreds of categories and sub-categories.
Moreover, the vector generation process is highly time consuming with processing abilities at the rate of 15-20 in one hour.

In summary, we report here a novel algorithm \sitevec\ for vector representation of binding sites which is fast, reliable and robust, suitable for high throughput binding site analysis in structual bioinformatics and drug discovery fields.


\section{Methods and materials:} \label{method}

    \sitevec\ is a mathematical enhancement to the PocketMatch \cite{yeturu2008pocketmatch} algorithm.
    In PocketMatch, pairwise distances are considered between representative points of a binding site.
    These distances are then put into bins corresponding to chemical types of pairs of site residues.
    The 20 amino acids are categorized into 5 groups – Group-0:(A,M,V,I,L,G,P); Group-1:(K,R,H); Group-2:(D,E,Q,N); Group-3:(Y,F,W); Group-4:(C,S,T) based on hydrophobic, hydrophilic and aromatic ring properties.
    Each amino acid of the site is represented by points - $C_\alpha, C_\beta$ and $C_\gamma$ for the centroid of the side-chain (including $C_\beta$).
    The PocketMatch algorithm takes as input {\it a pair of sites}.
    For a given site a variable dimensional descriptor is generated and for two sites, a similarity score is computed.
    As with any other pairwise method, limitations of PocketMatch include difficulty in serving proximity queries.
    For instance, if the query is to identify all sites that are proximal to a given novel nucleotide-binding site, it has to resort to several thousands of pairwise comparisons.
    The \sitevec\ algorithm attacks the descriptor variability issue in PocketMatch.
    Vector generation in \sitevec\ involves sliding window based processing of PocketMatch bins, clustering and auto-encoder based deep neural network derived embedding, the steps of which are described in the sections that follow.

    \subsection{Site2Vec: Sliding window-based encoding} \label{Site2Vecc}
    
\begin{figure*}
            \centering
            \centerline{\includegraphics[width=0.95\textheight, height=\textwidth, angle =90]{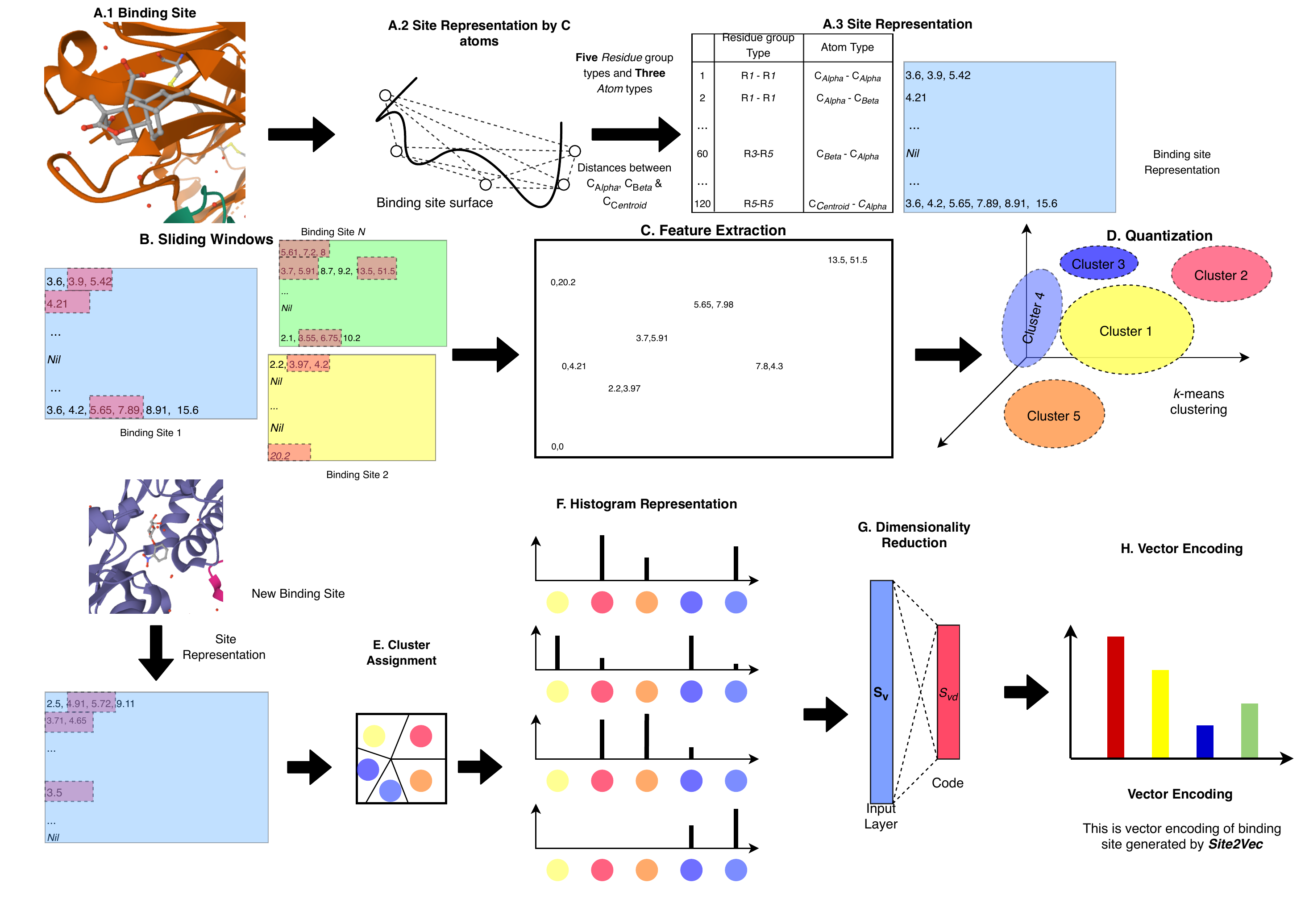}}
            \caption{Steps of Site2Vec for vector encoding of binding sites. Example of binding sites are in PDB id 1EFA \cite{bell2000closer} and 1KFA \cite{murata2002crystal}. The illusion is created by MOL* \cite{sehnal2018mol} and RCSB PDB \cite{berman2002protein}.}
            \label{fig:method}
        \end{figure*}

    \sitevec\ takes as input 3D representation of a binding site and encodes it into a $d$- dimensional vector representation (Figure~\ref{fig:method}).

        \begin{enumerate}
            \item \label{itm:1}\textbf{Representation of atoms in a binding site}
                \begin{center}

                    \begin{itemize}
                        \item[] Let $e$ be a binding site in dataset $\Gamma$ and $A(e)$ be the atoms of $e$.
                        \item[] $A(e)$ =$\{(x_i,y_i,z_i,\phi_i,\psi_i)|i \in [1..N]\}$
                        \item[] where,
                        \begin{itemize}
                            \item[] $x_i,y_i,z_i$ :3D coordinates of $A(e)$.
                            \item[] $\phi_i \in \{ \alpha,\beta,\gamma\}$.
                            \begin{itemize}
                                 \item[] $\alpha : C_\alpha$
                                 \item[] $\beta : C_\beta $
                                 \item[] $\gamma$ : Centroid of the side chain.
                            \end{itemize}
                            \item[] $\psi_i \in [1..k]$
                            \begin{itemize}
                                \item[] where, $k$ = 5 \cite{yeturu2008pocketmatch} 20 amino acids were considered in 5 groups – Group-0:(A,M,V,I,L,G,P); Group-1:(K,R,H); Group-2:(D,E,Q,N); Group-3:(Y,F,W); Group-4:(C,S,T).
                            \end{itemize}
                            \item[] $N$ is number of atoms
                        \end{itemize}

                    \end{itemize}
                \end{center}
            \item \label{itm:2} \textbf{Calculation of pairwise distances among the atoms of binding site}
                \begin{center}
                    \begin{itemize}
                        \item[] $D(e)$ = \{($dist(i,j),type(i,j)| \forall i,j \in[1..N]$\}
                        \item[] where,
                        \begin{itemize}
                            \item[]  $dist(i,j):\sqrt{\sum\limits_{p\in\{x,y,z\}}(A(e)[i].p-A(e)[j].p)^2}$
                            \item[] $type(i,j):A(e)[i].\phi\_A(e)[j]\phi\_A(e)[i].\psi\_A(e)[j].\psi$
                            \item[] such that,
                            \begin{itemize}
                                \item [] $\alpha\_\beta \equiv \beta\_\alpha$
                                \item [] $\alpha\_\gamma \equiv \gamma\_\alpha$
                                \item [] $\beta\_\gamma \equiv \gamma\_\beta$
                                \item [] $\psi_a\_\psi_b \equiv \psi_b\_\psi_a (\forall \psi_a \neq \psi_b)$
                            \end{itemize}
                        \end{itemize}
                    \end{itemize}
                \end{center}
            \item \label{itm:3} \textbf{Grouping of distances }
                \begin{center}
                    
                    \begin{itemize}
                        \item[] Let $|m|$ be the number of groups. 
                        \item[] $\forall m \in \{x[1]|x\in D\}$
                        \begin{itemize}
                            \item[] $L_m(e)=[x[0]|x\in D,x[1]=m] \land (\forall a<b) \rightarrow L_m[a]\leq L_m[b] $
                        \end{itemize}
                        
                    \end{itemize}
                \end{center}

            \item \label{itm:4} \textbf{Sliding a sliding window of size $\xi$ over the group of distances (calculated in the previous step)}
                \begin{center}
                    \begin{itemize}
                        \item[] $S_m^\xi(e)= [ L_m(e)[i:i+\xi]| 1\leq i \leq |L_m(e)|-\xi+1]$ 
                    \end{itemize}
                \end{center}
            \item \label{itm:5} \textbf{Creating a bag of features.}
                \begin{center}
                    \begin{itemize}
                        \item[] $S_m=\{S_m^\xi(e) | e \in \Gamma (Dataset)\}$ 
                    \end{itemize}
                \end{center}
            \item \label{itm:6} \textbf{Quantization}
                \begin{center}
                    \begin{itemize}
                        \item[] $O_m = Kmeans(S_m,k)$  (where $k$ is the number of clusters)
                    \end{itemize}
                    
                \end{center}
            \item \label{itm:7} \textbf{Mapping of each list (groups) to a histogram}
                \begin{center}
                    \begin{itemize}
                        \item[] $\forall e \in \Gamma$
                        \begin{itemize}
                            \item[] $C_m(e) = [O_m.nearest(s)|s \in S_m^\xi(e)]$ 
                            \item[] $H_m(e)= hist(C_m(e))$
                            \item[] where,
                            \begin{itemize}
                                \item[] $nearest(.)$ returns nearest cluster number \cite{beis1997shape}.
                                \item[] $hist(.)$ function generates histogram of cluster numbers.
                            \end{itemize}
                        
                        \end{itemize}
                    \end{itemize}
                \end{center}
            \item \label{itm:8} \textbf{Feeding histogram to an auto-encoder \cite{guo2016deep}, results in $d$ dimensional vector encoding}
            \begin{center}
                \begin{itemize}
                    \item[] $S_{m,d}^{\xi}(e) = \Psi(H_m(e),\theta^{*}) $
                    \item[] where,
                    \begin{itemize}
                        \item[] $\Psi$: is an auto-encoder
                        \item[] $\theta^{*} : \arg\min\limits_{\theta} \sum\limits_{e \in \Gamma} \|\Psi(H_m(e))-H_m(e)\|^2$
                        \item[] $S_{m,d}^{\xi}(e)$: $d$-dimensional vector representation of binding site $e$ using auto encoder $\Psi$.
                    \end{itemize}
                \end{itemize}
	            
	        \end{center}

        \end{enumerate}

	   \subsubsection{Training process} \label{training}
            
            In \sitevec, a sliding window (refer step \ref{itm:4}) passes over the list representation of binding sites (refer step  \ref{itm:3}) that generates equal length slices of distances of $ C_\alpha$, $C_\beta$ and $\gamma$ (refer step \ref{itm:4}) and the slices of distances are mapped to $k$ clusters such that each slice maps exactly one cluster (refer step \ref{itm:7}). 
            The default setting of this practice includes the size of sliding window $\xi=10$ (refer step \ref{itm:4}) and the number of clusters $k=10$ (refer step \ref{itm:6}).
            The default dimension of vector descriptor of binding site is $d=200$ (refer step \ref{itm:8}) where a protein-ligand binding site is represented as $m$ number of lists where $m=120$ \cite{yeturu2008pocketmatch} (refer step \ref{itm:4}).
            
            A 1D auto-encoder deep neural network is trained that results in $d$ dimensional vector encoding of binding sites (refer step \ref{itm:8}).
            The auto-encoder consists of four hidden layers, so a total of six layers, including input and output layers followed by the ReLU activation function.
            The configuration of the auto-encoder is $1200\times1000\times800\times600\times400\times200 $. The network is trained on a batch size of 16 with a learning rate of 0.0001.
            The setup mentioned above is the default configuration used in the study.
            
    \subsection{Simple mean-variance based vector representation}   
    
            In the previous section \ref{Site2Vecc}, we have mentioned that each binding site is represented as a $d$- dimensional vector. 
            In an intermediate step (step \ref{itm:3}) in section \ref{Site2Vecc}, a site is described a $m$ groups of distances between atoms.
            The first simple approach is to compute mean and variance for each of the groups mentioned in step \ref{itm:3} and generate a $m$-dimensional vector.
            The pseudo-code of this mean-variance based method is described in Algorithm \ref{alg:1}.
            \begin{algorithm}
                \renewcommand{\algorithmicrequire}{\textbf{Input:}}
                \renewcommand{\algorithmicensure}{\textbf{Output:}}
                \caption{Mean-Variance Method}
                \label{alg:1}
                \begin{algorithmic}[1]
                    \REQUIRE $e$ be a binding site.
                    \ENSURE $e_{v}$: Vector representation of the binding site.
                    \STATE $ e_{\mu} =[MEAN(L_i(e))]_{i=1}^{|m|}$ ($L_i$, same as step \ref{itm:3} in section \ref{Site2Vecc})
                    \STATE $ e_{\sigma}=[VARIANCE(L_i(e))]_{i=1}^{|m|}$ ($L_i$, same as step \ref{itm:3} in section \ref{Site2Vecc})
                    \STATE $ e_v = e_{\mu}\cdot e_{\sigma}$
                \end{algorithmic}
            \end{algorithm}

    \subsection{Uniform-Site2Vec: Uniform distribution of centroids based binding site encoding method }
    
        In section \ref{Site2Vecc}, \sitevec\ method is discussed and in step \ref{itm:6} , $K-means$ clustering algorithm \cite{jain1988algorithms} is used for grouping of $S_m$ (refer step \ref{itm:5} in \ref{Site2Vecc}) in $k$ clusters.
        Another simple and time-effective approach is to generate uniformly $k$ pseudo centers in $\xi$ space that represent $k$ clusters.
        The rest of the steps as well as the training process (\ref{training}) are same as in \ref{Site2Vecc}.
        This version of the method we call it {\it Uniform-Site2Vec}.
        
    \subsection{Discretized distance and histogram-based binding site representation method }
    
        In this approach, distances between $C_\alpha$, $C_\beta$ and $\gamma$ (Step \ref{itm:2}) are discretized into seven intervals (Table~\ref{Tab:intervalTable}).
        
            \begin{table}[!h]
            \centering
            \caption{ Discretization of distances between atoms into seven intervals \cite{weill2010alignment} }
            \begin{tabular}{@{}ll@{}}\toprule 
            \textbf{Interval} & \textbf{Distance range(in \AA)} \\
            \midrule
            1 & [0, 7.6) \\
            2 & [7.6, 10.1) \\
            3 & [10.1, 12.3) \\
            4 & [12.3, 14.3) \\
            5 & [14.3, 16.8) \\
            6 & [16.8, 20.0) \\
            7 & [20.0, $\infty)$\\
            \bottomrule
            \end{tabular}
            \label{Tab:intervalTable}
        \end{table}
       
       The distance elements within each PocketMatch based bin (refer step \ref{itm:2} and \ref{itm:3})) are mapped to these intervals and histogram of counts of number of elements in each interval (Table~\ref{Tab:intervalTable}) are computed.
       And finally, for each group ($L_m(e)$), the histogram of the interval number is generated.
       Concatenation of these histograms for all the groups is considered a vector representation (Algorithm~\ref{alg:2}).
       
        \begin{algorithm}
                \renewcommand{\algorithmicrequire}{\textbf{Input:}}
                \renewcommand{\algorithmicensure}{\textbf{Output:}}
                \caption{Discretized distance and histogram-based method}
                \label{alg:2}
                \begin{algorithmic}[1]
                    \REQUIRE $e$ be a binding site.
                    \ENSURE $e_{v}^{\textbf{B}}$: Vector representation of binding site.
                    
                    
                    \STATE $\textbf{B} = [\beta_1,\beta_2,\ldots,\beta_7]$
                    \STATE $X_{i}(e) = [\beta | \exists \beta \in$ \textbf{B}$: \beta_{k} \leq L_i(e)[j] < \beta_{k+1},$ where $k \in (1,6) ]_{j=1}^{|L_i(e)|} $
                    \STATE $X(e)=[X_i(e)]_{i=1}^{|m|}$
                    \STATE $e_v^{\textbf{B}} = [hist(X_i(e))]_{i=1}^{|m|}$ \\($hist(.)$ returns histogram of bin numbers.)
                \end{algorithmic}
            \end{algorithm}

    \subsection{Benchmarking Datasets and Methods}
     
    \sitevec\ algorithm has been evaluated on 10 different datasets and 23 existing methods for binding site comparison.
    The first study involves benchmarking of \sitevec\  against 21 methods and 7 datasets reported in \cite{ehrt2018benchmark}.
    The second study involves the evaluation of the algorithm on ApocS3 Dataset\cite{gao2013apoc}.
    The third study involves the evaluation of \sitevec\  on a data set of 2249 binding sites covering a class of 5 types of sites \cite{anand2014plic}.
    The fourth study involves the evaluation of the algorithm against another deep-learning based method in the literature \cite{pu2019deepdrug3d}.
    
    \subsubsection{Benchmarking on ProSPECCTs datasets and methods}
     
        \sitevec\ has been evaluated against 21 methods on 7 datasets based on \cite{ehrt2018benchmark}.
        
        Each of the methods are described in (Table~\ref{tab:literature}).
        The datasets are described in (Table~\ref{tab:prosecct-dataset-descriptions}).
        
        \begin{table}[]
            \centering
            \caption{Description of data sets used in \cite{ehrt2018benchmark} for large scale evaluation of methods for binding site comparison}
            \begin{adjustbox}{max width=\textwidth}
                \begin{tabular}{c|l} \toprule
                     \textbf{Data set} &  \textbf{Description}\\ \midrule
                     D1 & Structures having identical sequences\\
                     D1.2 & Structures with identical sequences and similar ligands\\
                     D2 & NMR structures. \\ 
                     D3 & Discrimination between binding sites having different physicochemical properties \\ 
                     D4 & Discrimination between binding sites having different physicochemical and geometrical properties \\ 
                     D5 & Kahraman dataset \cite{kahraman2010diversity} \\ 
                     D6 & Barelier dataset \cite{barelier2015recognition} \\ 
                     D7 & Binding site within known set of proteins \\ \bottomrule
                \end{tabular}
            \end{adjustbox}
            \label{tab:prosecct-dataset-descriptions}
        \end{table}

    \subsubsection{PLIC: protein-ligand interaction clusters}
        
        PLIC stands for Protein-ligand interaction clusters.
        It is a relational database that categorizes the ligand-binding sites based on different attributes \cite{anand2014plic}.
        The database mentioned above contains 67,550 unique protein-ligand binding sites which are divided into 10,854 groups based on pocket shape, types of residues, atomic contacts, and binding energy.
        Among these binding site groups, five groups that have more than 300 binding sites have been extracted for evaluating this method as any machine learning model requires a sufficient amount of data to learn each class properly. 
        The first group has 537 binding sites. The second, third, fourth, and fifth have 513, 473, 384 and 342 binding sites respectively.
    
    \subsubsection{ApocS3 Dataset}
        
        ApocS3 dataset is composed of subject and control data which consists of sites from 2090 and 21660 unique protein chains from PDB, respectively \cite{gao2013apoc}.
        There are 38,066 pairs of binding sites in subject set (positive pairs), and 38,066 pairs of binding sites in the control set (negative pairs).
        Subject set corresponds to known similar sites and control set corresponds to known dissimilar sites.
        
    \subsubsection{Comparison against nucleotide and heme binding sites}
        TOUGH-C1 dataset compiled by \cite{DVN/VMXOCT_2018} consists of nucleotide and heme binding sites.
        This data set consists of 1556 nucleotide pockets and 556 heme- binding sites.
        We have used this data set for benchmarking of \sitevec\ with respect to \cite{pu2019deepdrug3d} which is another recent study on using convolutional neural networks for binding site classification.

    \subsection{Evaluation Technique}
\label{evlatech}
    A two class clarification algorithm can be evaluated on standard metrics such as precision, recall, accuracy and AUC \cite{YETURU202081}.
    AUC stands for Area Under the Curve of a graph plotted against true positive and false positive rates of prediction of a classifier at various thresholds.
    The performance of binding site comparison and classification methods is represented by the receiver operating characteristic (ROC) curve \cite{bradley1997use}.
    Area Under the Curve (AUC) of ROC is a standard quality assessment metric.
    ROC curve plots True Positive Rate (TPR) with respect to False Positive Rate (FPR) of a classifier about several threshold values on its confidence score.
    In the ROC curve, TPR is plotted on the Y-axis and FPR is plotted on the X-axis. 
    The TPR and FPR are defined below.
    $$TPR = \frac{TP}{TP+FN}$$ and $$FPR=\frac{FP}{FP+TN}$$
    Besides the ROC curve, there are other measuring tools like Accuracy, Precision and Recall that are also used to evaluate the performance of protein-ligand binding site comparison methods. 
    These metrics are described as \\
    $$ Accuracy = \frac{TP+TN}{TP+FP+TN+FN}$$
    $$Precision = \frac{TP}{TP+FP}$$
    $$Recall = \frac{TP}{TP+FN}$$
    where TP is the number of true positives, FP is the number of false positives, TN is the number of true negatives and FN is the number of false negatives.
    In this exercise of benchmarking against other methods \cite{weill2010alignment, pu2019deepdrug3d, nakamura2016protein} we have used AUC as a quality assessment metric.
    The state of the art methods given in \cite{ehrt2018benchmark} provide a similarity score between a pair of sites which is used to compute ROC curve.
    
    \subsection{Run time evaluation}
\label{runtime}

    In this study, we have computed the run time performance of \sitevec and compared it against \cite{pu2019deepdrug3d}.
    The run time is computed on an Intel Xeon workstation, having CPU E5-2640 and with 2.40 GHz, 40 cores and 125GB RAM and Ubuntu 16.04.5 LTS is installed.
    The site comparisons have happened in a serial manner.

\section{Results:}

    \sitevec\ algorithm has been evaluated on 10 different datasets and several methods for binding site comparison.
    The first study involves benchmarking of the method against 21 methods and 7 datasets reported in \cite{ehrt2018benchmark}.
    The second study involves evaluation of the algorithm on ApocS3 dataset\cite{gao2013apoc}.
    The third study involves the evaluation of the method on a data set of 67550 binding sites covering a class of 5 types of sites \cite{anand2014plic}.
    
    The fourth study involves the evaluation of the algorithm against another deep-learning based method in the literature \cite{pu2019deepdrug3d}.
    
    \subsection{Benchmarking against existing binding site comparison methods on ProSPECCTs datasets}

        The \sitevec\ model is trained on whole PLIC dataset \cite{anand2014plic} and represents a binding site as a $200$-dimensional vector.
        For each of the ProSPECCTs Datasets \cite{ehrt2018benchmark}, \sitevec\ vector is generated for each site.
        For a pair of sites, the corresponding vectors are fused to result in a concatenated vector.
        The fused vector is then used in classification tasks.
        For finding similarity, each dataset is divided into two halves randomly. A data-driven classifier is trained by one half.
        The classifier is tested by the remaining half of the data and AUC of ROC curve measures performance.
        Our approach is compared to 21 different binding site comparison methods mentioned in ProSPECCTs.
        For all the previous methods, the AUC scores are directly obtained from \cite{ehrt2018benchmark} without having to run the methods locally again.
        Given that \sitevec\ is a vector embedding generator, it needs to have train and test sets for application in a classification scenario.
        To this end, we have considered each data set provided in \cite{ehrt2018benchmark} and split it into train and test sets.
        A classifier (Random Forest \cite{liaw2002classification} having 100 trees) is trained on the train set and evaluated for AUC value on the test set.
        Instead of single random split of train and test sets, we have randomly split the data set into two halves 50 times and computed the average AUC value.

        Dataset D1 consists of structures having identical sequences that bind to different ligands.
        It has 13,430 similar pairs and 92,846 dissimilar pairs.
        Dataset D1.2 evaluates sensitivity to the binding sites with identical sequences and similar ligands.
        Our proposed approach achieved AUC scores of 1 and 0.94 on D1 and D1.2, respectively.
        Probis \cite{konc2010probis}, SMAP \cite{xie2009unified}, TM-align \cite{zhang2005tm} have also performed very well and got AUC of 1 on both datasets.
        From the AUC scores of 1 and AUC 0.94, we can conclude that \sitevec\ is robust method for the structural definition of binding sites.
    
        Dataset D2 measures the sensitivity with respect to the pocket flexibility of Nuclear Magnetic Resonance (NMR) structures.
        \sitevec\ has scored AUC of 1 which concludes that it is sensitive to binding site flexibility.
    
        Two decoy datasets D3 \& D4 are compiled to measure the difference between nearly similar sites differing by physicochemical properties (D3) and both physicochemical and geometrical properties by performing five mutations as mentioned in ProSPECCTs\cite{ehrt2018benchmark}.
        \sitevec\ scores better (AUC 0.99) than the other 21 methods.
        Hence, we conclude that our \sitevec\ is robust in handling mutation as well.
    
        Next, dataset D5 is compiled by \cite{kahraman2010diversity} and has 10000 pairs of binding sites for comparison. 
        On this data set, \sitevec\ achieves an AUC of 0.86 which is highest among other methods (Table~\ref{Tab:proSPECCT}).
        Dataset D6, compiled by \cite{barelier2015recognition}, comprises of protein-ligand binding sites with identical ligands.
        On dataset D6, using default Random Forest classifier with 100 trees and unlimited depth, our approach performs moderately by scoring AUC of 0.53 where KRIPO \cite{wood2012pharmacophore} outperforms all other methods, including \sitevec\ with an AUC of 0.73.
        However using Gradient Boosting Classifier on this data set, we obtain an AUC value of about 0.75.
        These experiments clearly indicate that by doing a hyper parameter grid search, \sitevec\ vectors can be used to improve classifier accuracy and AUC metrics.

        Lastly, dataset D7 is compiled to measure the recovery of known binding site similarities within a set of diverse proteins.
        There are only 115 positives and 56284 negatives in this data set.
        It has heavy bias towards negative data set.
        \sitevec\ achieves an AUC of 0.66, which is lower than most pocket matching algorithms when default Random forest classifier is used.
        However, when Ada Boost Classifier is used with 100 trees, the AUC value boosts up to 0.80.
        We can therefore infer that \sitevec\ vectors can be considered along with machine learning model hyper parameters to result in high quality scores.

        On most datasets, \sitevec is outperforming other existing methods.
        The benchmarking study supports the claim that vectorization approach is robust and reliable vectors are generated out of binding sites in \sitevec.
        In cases where a Random Forest classifier was not suitable, an alternative ensembling method such as Ada Boost or Gradient Boost methods with appropriate parameter settings have resulted in high AUC scores.
        This exercise infers that \sitevec\ vectors are of good quality and can be used in classification scenarios with appropriate machine learning model selection through hyper parameter search.

    \begin{table}[!h]
        \centering
        \caption{AUC values for 21 existing site similarity methods and Site2Vec on ProSPECCTs datasets}
        \begin{adjustbox}{max width=\textwidth}
        \begin{tabular}{@{}l|lllllllll@{}}\toprule 
                \textbf{Method} &\textbf{D1} & \textbf{D1.2} & \textbf{D2} & \textbf{D3} & \textbf{D4} & \textbf{D5} & \textbf{D6} & \textbf{D7} &    \begin{tabular}{cc}
                     \small{Average AUC}\\
                     \small{across data sets}
                \end{tabular}\\\midrule
                \textbf{Site2Vec*} & 1.00 & 0.94 & 1.00 & 0.99 & 0.99 & 0.86 & 0.53 & 0.66 & 0.87 \\ 
                \textbf{SiteHopper} \cite{batista2014sitehopper} & 0.98 & 0.94 & 1.00 & 0.75 & 0.75 & 0.81 & 0.56 & 0.77 & 0.82 \\
                \textbf{SiteEngine} \cite{shulman2005siteengines} & 0.96 & 1.00 & 1.00 & 0.82 & 0.79 & 0.57 & 0.55 & 0.86 & 0.82 \\
                \textbf{SiteAlign} \cite{schalon2008simple} & 0.97 & 1.00 & 1.00 & 0.85 & 0.8 & 0.57 & 0.44 & 0.87 & 0.81 \\
                \textbf{SMAP} \cite{xie2009unified} & 1.00 & 1.00 & 1.00 & 0.76 & 0.65 & 0.54 & 0.68 & 0.86 & 0.81 \\
                \textbf{KRIPO } \cite{wood2012pharmacophore} & 0.91 & 1.00 & 0.96 & 0.6 & 0.61 & 0.77 & 0.73 & 0.85 & 0.80 \\
                \textbf{Shaper} \cite{desaphy2012comparison} & 0.96 & 0.93 & 0.93 & 0.71 & 0.76 & 0.65 & 0.54 & 0.75 & 0.78 \\
                \textbf{Shaper(PDB)} \cite{desaphy2012comparison} & 0.96 & 0.93 & 0.93 & 0.71 & 0.76 & 0.64 & 0.54 & 0.75 & 0.78 \\
                \textbf{VolSite/Shaper} \cite{desaphy2012comparison} & 0.93 & 0.99 & 0.78 & 0.68 & 0.76 & 0.58 & 0.71 & 0.77 & 0.78 \\
                \textbf{FuzCav} \cite{weill2010alignment} & 0.94 & 0.99 & 0.99 & 0.69 & 0.58 & 0.54 & 0.67 & 0.77 & 0.78 \\
                \textbf{FuzCav(PDB)} \cite{weill2010alignment} & 0.94 & 0.99 & 0.98 & 0.69 & 0.58 & 0.54 & 0.65 & 0.77 & 0.77 \\
                \textbf{TM-align} \cite{zhang2005tm} & 1.00 & 1.00 & 1.00 & 0.49 & 0.49 & 0.62 & 0.59 & 0.88 & 0.76 \\
                \textbf{Cavbase} \cite{schmitt2002new} & 0.98 & 0.91 & 0.87 & 0.65 & 0.64 & 0.57 & 0.55 & 0.82 & 0.75 \\
                \textbf{VolSite/Shaper(PDB)} \cite{desaphy2012comparison} & 0.94 & 1.00 & 0.76 & 0.68 & 0.76 & 0.56 & 0.5 & 0.72 & 0.74 \\
                \textbf{IsoMIF} \cite{chartier2015detection} & 0.77 & 0.97 & 0.7 & 0.59 & 0.59 & 0.81 & 0.62 & 0.87 & 0.74 \\
                \textbf{PocketMatch} \cite{yeturu2008pocketmatch} & 0.82 & 0.98 & 0.96 & 0.59 & 0.57 & 0.6 & 0.51 & 0.82 & 0.73 \\
                \textbf{ProBiS} \cite{konc2010probis} & 1.00 & 1.00 & 1.00 & 0.47 & 0.46 & 0.55 & 0.5 & 0.85 & 0.73 \\
                \textbf{TIFP} \cite{desaphy2013encoding} & 0.66 & 0.9 & 0.91 & 0.66 & 0.66 & 0.63 & 0.55 & 0.71 & 0.71 \\
                \textbf{RAMPAD} \cite{krotzky2015large} & 0.85 & 0.83 & 0.82 & 0.61 & 0.63 & 0.52 & 0.6 & 0.74 & 0.70 \\
                \textbf{Grim} \cite{desaphy2013encoding} & 0.69 & 0.97 & 0.92 & 0.55 & 0.56 & 0.61 & 0.45 & 0.7 & 0.69 \\
                \textbf{Grim(PDB)} \cite{desaphy2013encoding} & 0.62 & 0.83 & 0.85 & 0.57 & 0.56 & 0.58 & 0.45 & 0.64 & 0.64 \\
                \textbf{TIFP(PDB)} \cite{desaphy2013encoding} & 0.55 & 0.74 & 0.78 & 0.56 & 0.57 & 0.53 & 0.56 & 0.66 & 0.62\\\bottomrule
            
            \end{tabular}
        \end{adjustbox}
        \label{Tab:proSPECCT} 
    \end{table}

        \subsection{Evaluation of quality of Vector representation on APOCS3 Dataset}
    
        ApocS3 dataset \cite{gao2013apoc} consists of 38,066 pairs of binding sides in subject (positive pairs) as well as 38,066 pairs of sites in control (negative pairs) set. 
        The positive data denotes known set of similar sites and the negative data denotes known set of dissimilar sites.
        
        \begin{figure}
            \begin{center}
            \includegraphics[scale=0.5]{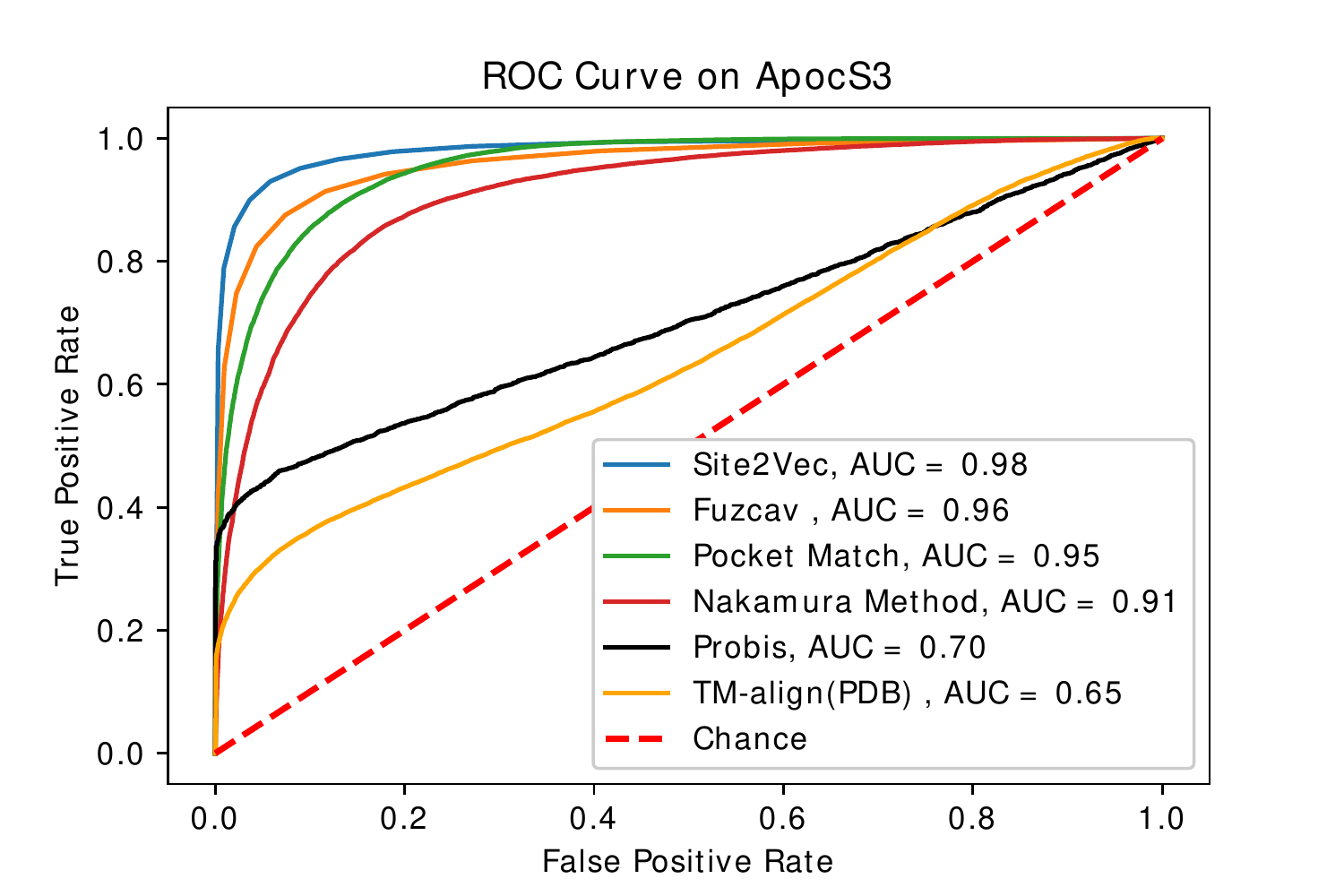}
            \caption{ROC plot and its corresponding AUC values measuring the performance of different methods and our method on APOCS3 dataset \cite{gao2013apoc}. The x-axis is false positive rate, and the y-axis is true positive rate.}
            \label{fig:apos}                
            \end{center}

        \end{figure}
        Using \sitevec, vector fingerprints of all 23,750 binding sites are generated.
        To find a similarity between two sites, the whole dataset (Subject and Control dataset) is divided into two halves.
        One half is used as a training data, and the other half is treated as test data such that both train and test sets have an equivalent amount of subject and control data.
        The pair of binding sites in the training data are fed to a classifier (Random Forest classifier containing 1000 trees) to train the similarity between two ligand-binding sites, and test data is used to predict the similarity score.
        The performance is measured by Receiver Operating Characteristic (ROC) curve and its corresponding area under the curve (AUC) score.

        We compared our proposed method with alignment dependent methods (i) TM-align \cite{zhang2005tm} and Probis \cite{konc2010probis} and (ii) alignment independent methods - Fuzcav \cite{weill2010alignment}, method by Nakumara and co-workers \cite{nakamura2016protein} and PocketMatch \cite{yeturu2008pocketmatch}.
        Brief description of these methods can be found in (Table~\ref{tab:literature}).
        
        The measured ROC curves are displayed in (Figure~\ref{fig:apos}).
        Our proposed method achieved an area under the curve of $0.983$ and is outperforming all other methods discussed in this study.
        FuzCav (AUC $0.96$), PocketMatch (AUC of $0.95$), Nakamura method (AUC of $0.91$) are also performing better on the ApocS3 dataset.
        Alignment dependent binding site similarity methods Probis has AUC of $0.70$ and TM-align achieves AUC of $0.65$.
        This result provides an insight that the quality of vector encoding of binding sites computed by \sitevec\ is acceptable and it is noted that
        our method is capable of doing multiple comparisons of binding sites at a single point of time. The proposed method converts binding sites to vectors in $d-$ dimensional space and compares vector representations in that multidimensional space.

        \begin{table}[!h]
        \centering
        \caption{Performance measure of Site2Vec and other binding site comparison methods on the ApocS3 dataset.}
        \begin{tabular}{@{}lllll@{}}\toprule \textbf{Method} & \textbf{Accuracy(Threshold)} & \textbf{AUC} &\textbf{Precision} & \textbf{Recall} \\\midrule
        \sitevec & \textbf{0.93(0.5)} & \textbf{0.98} & \textbf{0.94} & \textbf{0.93}\\
        FuzCav & 0.90(0.5) & 0.96 & 0.88 & 0.91 \\
        PocketMatch & 0.86(0.1) & 0.95 & 0.92 & 0.79 \\
        Nakamura method & 0.84(0.5) & 0.91 & 0.80 & 0.89\\
        Probis & 0.61(0)& 0.70 & 0.90 & 0.59\\
        TM-align & 0.62(0.4) & 0.65 & 0.89 & 0.28 \\\bottomrule

        \end{tabular}
        \label{Tab:03}
        
        \end{table}

        \subsection{Benchmarking against DeepDrug3D}
    
        Another method available today that employs deep neural networks for binding site embedding is DeepDrug3D \cite{pu2019deepdrug3d}.
        This is a classification algorithm that uses convolutional neural networks (CNN) to classify a given data set of categories of binding sites.
        In the process, it generates a vector embedding once the training is complete for a given binding site.
        In this method, a binding site is represented as a voxel representation with a 3D grid.
        For each grid point, interaction energy is calculated for 14 atom types. 
        The voxel-grid is then passed through a convolutional neural network which regresses over a loss function based on category labels of the input sites.
        As with any CNN, the method is sensitive to input representation and reference frame.
        To address this, the authors use a pre-processing step where principal component analysis (PCA) is carried out to determine X, Y and Z axes and align the site and translate the origin to the centroid of the site.
        
        However, fundamental difference between DeepDrug3D and \sitevec\ is, the former is fundamentally a classification technique that requires categories of binding sites in the data set.
        \sitevec\ is a vector embedding technique that does not require categories of binding sites to generate vector representation.
        A plain vector embedding generator has clear advantages over fixed classifiers as it scales to several thousands of sites across diverse categories in dynamic pipeline processing scenarios.
        
        The first benchmarking we did was on PLIC \cite{anand2014plic} data set for both AUC scores and the second exercise was on TOUGH-C1 data set\cite{DVN/VMXOCT_2018}.
        We can very clearly see that \sitevec\ is faster at scale (about 3400 times) than DeepDrug3D and performs better or at par with the latter method.
        
        \subsubsection{Performance on PLIC data set}
        
\begin{figure}[!t]
            \begin{center}
            \includegraphics[scale=0.5]{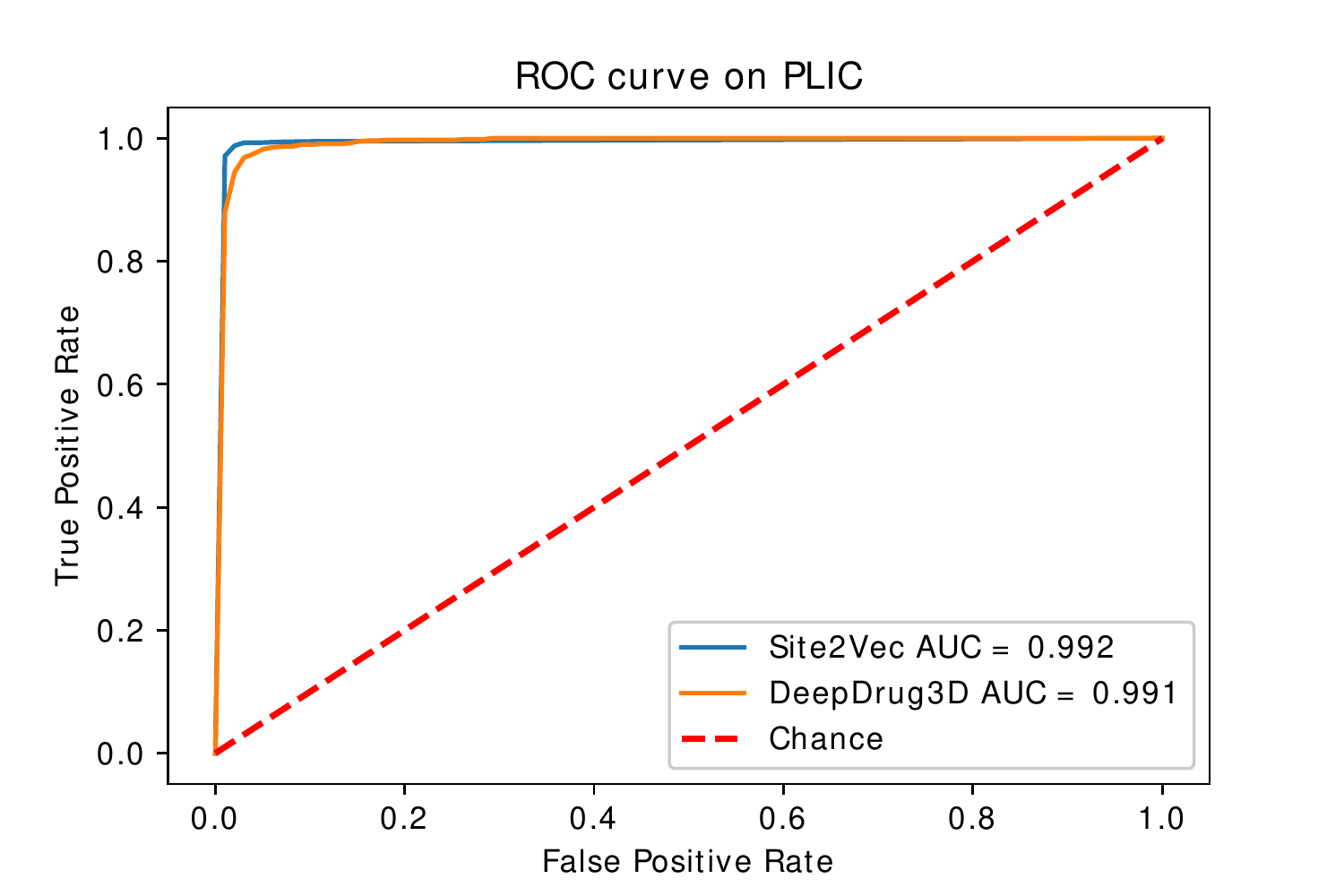}
            \caption{ROC plot for evaluating the proposed method and Deepdrug3D to classify the protein-ligand binding sites into five groups. The x-axis represents the False positive rate, and the y-axis denotes True positive rate.}
            \label{fig:rocProposedMethodPLIC}                
            \end{center}

        \end{figure}

        Comparison studies are carried out against another deep learning based method Deepdrug3D \cite{pu2019deepdrug3d} on the same data set.
        Figure{ \ref{fig:rocProposedMethodPLIC}} indicates the ROC curve for the PLIC dataset for both the methods.
        We can observe that both the methods have obtained very high AUC of above 0.99.
        
        In addition to a random forest classifier, other classifiers are also considered while performing the evaluation process.
        K-nearest neighbor classifier and Decision tree \cite{kotsiantis2007supervised} are used to assess the performance of vector representation of \sitevec.
        Table \ref{Tab:01} shows the accuracy, area under the curve of ROC plot of \sitevec\ using different classifiers and Deepdrug3D on the PLIC dataset.
        We can observe that \sitevec\ has slightly out performed \cite{pu2019deepdrug3d} on this data set.
        \begin{table}[!h]
            \centering
            \caption{Performance analysis of \sitevec\ and Deepdrug3D to classify the binding sites on PLIC data set.}
            \begin{tabular}{@{}lll@{}}\toprule \textbf{Method} & \textbf{Accuracy} & \textbf{AUC} \\\midrule
                K-NN on Site2Vec & 0.97 & 0.990\\
                Decision tree on Site2Vec & 0.95 & 0.976 \\
                Random Forest on Site2Vec & \textbf{0.98} & \textbf{0.992}\\
                DeepDrug3D & 0.96 & 0.991 \\\bottomrule
            \end{tabular}
            \label{Tab:01} 
        \end{table}
        The \sitevec\ algorithm comes with a very clear advantage of ultra fast execution speed compared with \cite{pu2019deepdrug3d}.
        It is at least {\bf \it 3400 times faster} for vector generation.
        The main reason why \cite{pu2019deepdrug3d} is slow compared to \sitevec\ is, it is a CNN based method which is fundamentally a classification approach.
        It requires full level of training to be able to generate a vector embedding.
        Whereas in \sitevec, absolute representation of site as in \cite{yeturu2008pocketmatch} is used followed by a normalization step of clustering and auto-encoder to generate generate embedding which is faster at scale.

    \subsubsection{Performance on TOUGH-C1}
    
    The TOUGH-C1 dataset \cite{DVN/VMXOCT_2018} has 4 sub-sets nucleotide-, heme-,steroid- and control- binding pockets.
    
\begin{figure}[h!]
            \begin{center}
            \includegraphics[scale=0.5]{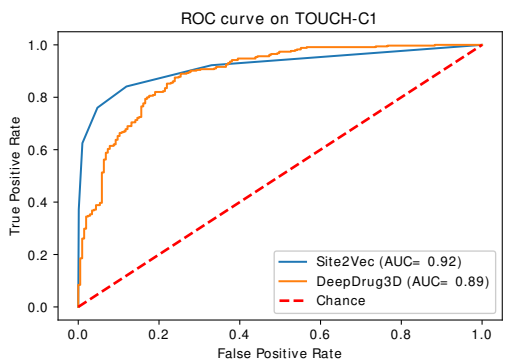}
            \caption{ROC plot for evaluating of \sitevec\ and Deepdrug3D to classify protein-ligand binding sites. The x-axis represents the False positive rate, and the y-axis denotes True positive rate.}
            \label{fig:siteDeepTouch}                
            \end{center}

        \end{figure}

    In this benchmarking exercise, heme-binding sites are treated as positive class data, and nucleotide sites are considered a negative class.
    The whole dataset of TOUGH-C1 is divided into 70\% of training data and 30\% test data.
    \sitevec\ is used to generate vector fingerprint from the binding site structure, and a classifier (Random Forest. Number of trees is 100) is used to classify protein-ligand binding sites using their vector form.
    The performance is measured using the ROC curve.
    The resultant ROC is shown in figure \ref{fig:siteDeepTouch}.
    \sitevec has got AUC of 0.92, and DeepDrug3D achieved an AUC of 0.89.
    Table \ref{Tab:deep2} gives more insight into the performance of \sitevec and Deepdrug3D on the TOUGH-C1 dataset from which we can infer that \sitevec\ has performed well heme and nucleotide binding sites of the TOUGH-C1 dataset.
    The TOUGH-C1 dataset \cite{DVN/VMXOCT_2018} has 2 other sub-data sets, called control and steroid sets on which \cite{pu2019deepdrug3d} has been tested.
    We chose to not evaluate against these sub-data sets as we felt it is not required in the context of a large scale evaluation already happened against \cite{ehrt2018benchmark}.
    In the \cite{ehrt2018benchmark}, \sitevec, has out performed several other algorithms and has been the top performing algorithm.
    Moreover, the amount of time requirement by DeepDrug3D \cite{pu2019deepdrug3d} takes several days to weeks on \cite{ehrt2018benchmark} benchmarking data sets to carry out any rigorous large scale study.

        \begin{table}[!h]
            \centering
            \caption{Performance measure of \sitevec\ and DeepDrug3D on the classification of nucleotide- and heme-binding sites.}
            \begin{adjustbox}{max width=\linewidth}
           \begin{tabular}{@{}lllll@{}}\toprule \textbf{Method} & \textbf{Accuracy} & \textbf{AUC} &\textbf{Precision} & \textbf{Recall} \\\midrule
            Site2Vec method & \textbf{0.90} & \textbf{0.92} & \textbf{0.88} & 0.71\\
            DeepDrug3D  & 0.84 & 0.89 & 0.85 & \textbf{0.89}\\
        \bottomrule
        \end{tabular}
        \label{Tab:deep2}
        \end{adjustbox}
        \end{table}

        \subsection{Comparison of variant of alternative vectorizations of \sitevec\ on TOUGH-C1 Dataset}
    
    The idea of this exercise is to assess different vectorization alternatives under \sitevec\ problem formulation.
    Once distance elements are available in 120 bins, vectorization can happen using any of the approaches as in section \label{method}.
    We have used TOUGH-C1 data set \cite{DVN/VMXOCT_2018} to compare within \sitevec\ formulation and decide on which vector generation variant is promising.
    We have used nucleotide and heme binding sites as two categories against which tested \sitevec\ internal variations.
    
    Here, heme-binding sites are treated as positive class data, and nucleotide sites are considered as negative class.
    The whole dataset is divided into 70\% of train data and 30\% of test data.
    From the binding site structures, methods described in section \ref{method}, are used to generate vector signatures and then classifiers are used to classify the sites using the vector representations of ligand-binding sites.
    Table \ref{Tab:deep} reports the performance of binding site vector encoding methods, and the performance is measured using accuracy and area under ROC. 
    Apart from the performance measure, the time required to encode binding sites by different proposed methods is also calculated and the average time taken by each method is listed in Table \ref{Tab:deep}.
    It is observed that \sitevec\ using PLIC \cite{anand2014plic} as clustering data set has out performed other variants.
    Based on this exercise we have set default \sitevec\ configuration to use clustering model for sliding window vectors from PLIC data set.
    
    \begin{table}[!t]
        \centering
        \caption{Performance measure of different encoding methods on the classification of nucleotide- and heme-binding sites.}
        \begin{tabular}{@{}llllll@{}}\toprule \textbf{Method} & \textbf{Accuracy} & \textbf{AUC} &\textbf{Precision} & \textbf{Recall} & \textbf{Avg. Time}\\\midrule
        Site2Vec method & \textbf{0.90} & \textbf{0.92} & \textbf{0.88} & \textbf{0.71} & 0.32 s\\
        Mean variance method  & 0.81 & 0.83 & 0.67 & 0.61 & 0.09 s\\
        Uniform-Site2Vec & 0.88 & 0.90 & 0.84 & \textbf{0.71} & 0.39 s\\
        Discretized distance  method & 0.86 & 0.90 & 0.87 & 0.62 & \textbf{0.08 s}\\
        \bottomrule
        \end{tabular}
        \label{Tab:deep}
        \end{table}

    \subsection{Run Time analyses}

    In this exercise we have estimated time of execution of the \sitevec\ algorithm and compared against other methods in the state of the art.
    Apart from site vector quality benchmarking exercises, 
    In this exercise, 30 structures were randomly chosen from a data set by Kahraman and co-workers\cite{kahraman2010diversity} and running times were calculated on \sitevec\, PocketMatch \cite{yeturu2008pocketmatch}, TM-align \cite{zhang2005tm} and Probis \cite{konc2010probis}.
    Brief descriptions of the above methods can be found in (Table ~\ref{tab:literature}).
    After performing this experiment (Table~\ref{Tab:time}), it is found that among all the methods studied here as well as Site2Vec, PocketMatch is the most computational cost-efficient method. 
    On average, PocketMatch takes 0.009s to compare pairs of binding sites where \sitevec\ consumes average 0.03s for comparison.
    TM-align and Probis compare two bind sites, on average 0.33s and 1.99s, respectively.
    We can observe that \sitevec\ is fastest in single site vector embedding generation methods in these methods and data set.
        \begin{table}[!h]
            \centering
            \caption{Run Time analyses of \sitevec\ and other binding site comparison algorithms for comparison binding sites}
            \begin{tabular}{@{}ll@{}}\toprule \textbf{Method} & \textbf{Avg. Comparison time(s)}\\\midrule
                Pocket Match & 0.009s \\
                Site2Vec* & 0.03s \\
                TM-Align & 0.33s\\
                Probis & 1.99s \\ \bottomrule
            \end{tabular}
            \label{Tab:time} 
        \end{table}

    Deepdrug3D \cite{pu2019deepdrug3d} follows a different approach for classifying the binding sites.
    It uses 3D voxel to represent a binding site and this voxel is fed into CNN for classification.
    Another benchmarking exercise is carried out to measure the time taken to convert binding sites into vector forms by Site2Vec and voxels form by Deedrug3D.
    In this study, twenty binding site is taken from \cite{DVN/VMXOCT_2018}.
    Deepdrug3D is computationally very expensive method with respect to run time. 
    It takes on an average of 1228.24s and \sitevec is taking 0.34s to encode a binding site.
    \sitevec\ is about 3400 times faster than \cite{pu2019deepdrug3d} on this data set.
    \begin{table}[!h]
        \centering
        \caption{Run Time analyses of Site2Vec and DeepDrug3D}
            \begin{tabular}{@{}ll@{}}\toprule \textbf{Method} & \textbf{Avg. run time(s)}\\\midrule
                Site2Vec* & 0.34s \\
                DeepDrug3D & 1228.24s\\\bottomrule
            \end{tabular}
            \label{Tab:timcomp} 
    \end{table}
    Both computational cost measure exercises are performed on Intel Xeon workstation, details mentioned in section \ref{runtime}.

\section{Discussion:}

    \subsection{Data visualization}

    \sitevec\ is an algorithm that generates vector embedding for a binding site.
    It is different from any pairwise comparison method which takes two sites as input and compute a score.
    The advantage of a pure vector embedding generator method is its usefulness as a locality sensitive hashing function.
    Proximity queries for binding sites in the vector embedded dimensional space would correspond to similar binding sites.
    To evaluate the quality of the vector representation, we have considered PLIC data set \cite{anand2014plic} and generated vectors for each of the sites.
    The data set has 5 categories of ligands.
    The 200 dimensional vectors hence generated using \sitevec\ are then projected onto a 2 dimensional space and plotted for visual inspection.
    We have used t-Distributed Stochastic Neighbor Embedding (t-SNE) \cite{maaten2008visualizing} for the projection purpose.
    
\begin{figure}[!h]
            \begin{center}
            \includegraphics[scale=0.5]{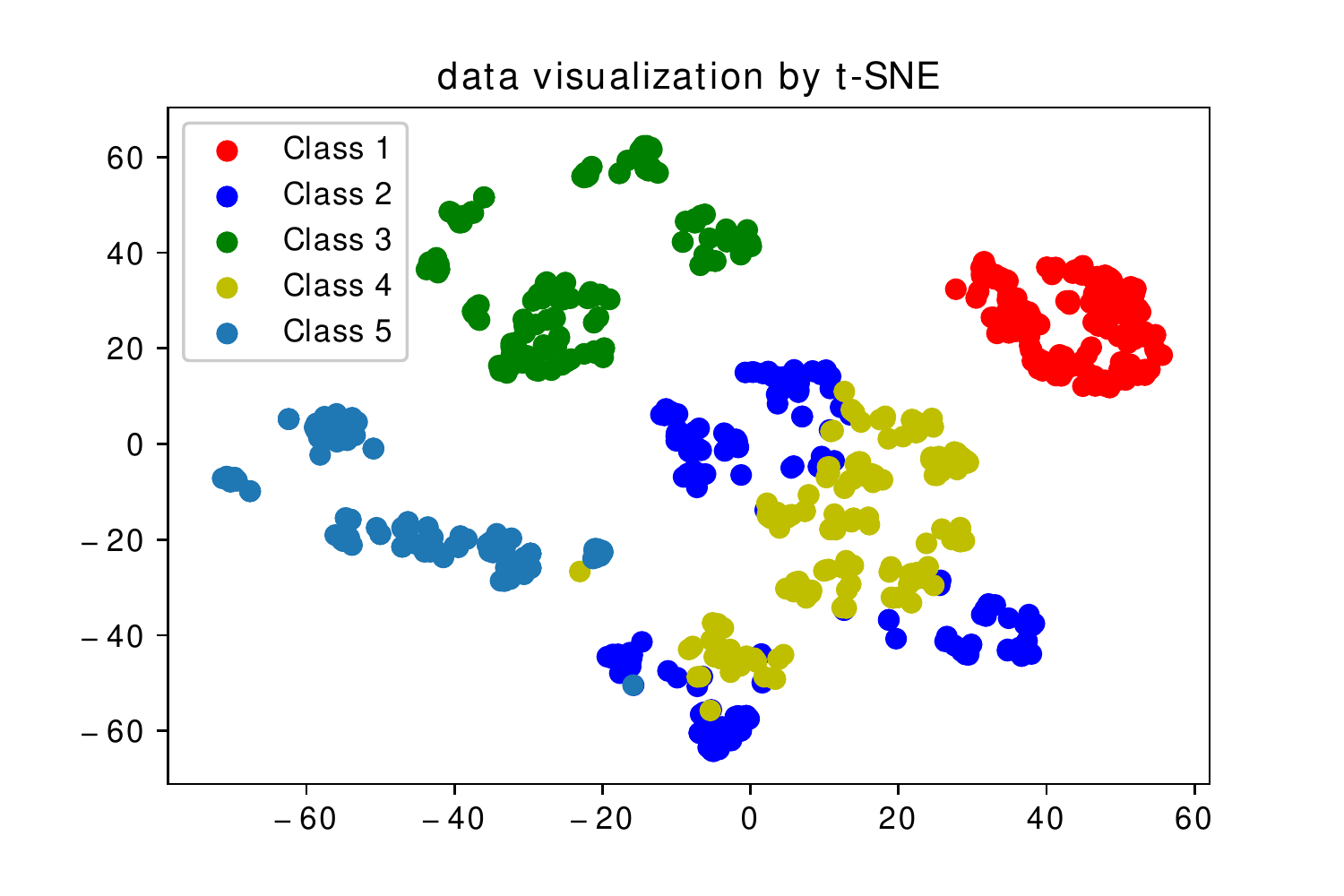}
            \caption{2D t-SNE visualization of vector embedding of binding sites of PLIC dataset labeled based on groups.}
            \label{fig:tsne}                
            \end{center}
        \end{figure}

    We can observe (Figure~\ref{fig:tsne}) that similar sites have occured in proximal locations.
    As can be seen from the colour of the points.
    Sites belonging in same class have same colour.
    This experiment supports the hypothesis that the vector embeddings generated by \sitevec\ capture spatial proximity well.
    It is the first novel locality sensitive hashing function in the state of the art in the space of binding site vectorization algorithms.
    
        \subsection{Metrics for comparison of binding site pair}
                
                Protein-ligand binding site comparison methods use different techniques to compare sites.
                The metric used for computing similarity is closely tied to the way data is represented.
                For instance alignment based method such as PocketAlign \cite{yeturu2011pocketalign} uses the number of mapped residues between sites and SitesBase \cite{gold2006sitesbase} uses matching atoms between sites.
                In case of alignment free methods, PocketMatch \cite{yeturu2008pocketmatch} uses number of matching distance elements between sites, RAPMAD \cite{krotzky2014large} uses the Jensen-Shannon divergence between histograms of distance elements between sites and FuzCav \cite{weill2010alignment} uses intersection of fingerprint vectors.
                
                In \sitevec\ method, cosine similarity (Equation~\ref{eqn:cosinesimi})  and euclidean distance (Equation~\ref{eqn:euclidean}) between vector embeddings are used metrics for similarity and dissimilarity respectively.
                Cosine similarity captures directionality of the sites where as Euclidean distance only looks at magnitude of separation between two sites and algebraically both are related quantities given angle between the two vectors.
                These metrics are used in sensitivity analysis of the method.
                
                \begin{equation}
                \label{eqn:cosinesimi}
                    cosine Similarity =\frac{\textbf{A.B}}{\parallel\textbf{A}\parallel\parallel\textbf{B}\parallel}\\
                 \> =\frac{\sum_{i=0}^{d-1}A_iB_i}{\sqrt{\sum_{i=0}^{d-1}A_i^2}\sqrt{\sum_{i=0}^{d-1}B_i^2}}
                     \end{equation}
                
                where, 
                
                \textbf{A} and \textbf{B}: vector representation of protein-ligand binding sites, 
                $d$ : The dimension of vector embedding of the binding site.\\
                The cosine similarity ranges between -1 and 1. 
                The value -1 implies that two vectors are opposite, and 1 indicates that the two vectors are the same and similarity value of 0 indicates orthogonality.
                The intermediate values indicate intermediate similarity. 
                
                 Euclidean distance between the pair of vector signatures of binding sites is used to calculate the similarity between two binding sites.
                \begin{equation}
                \label{eqn:euclidean}
                    e(\textbf{A},\textbf{B}) = \sqrt{\sum_{i=0}^{d-1}(A_i-B_i)^2}
                \end{equation}
                
                where,
                
                \textbf{A} and \textbf{B}: Vector signatures of two binding sites.
                
                $e$(\textbf{A},\textbf{B}): Euclidean distance between \textbf{A} and \textbf{B}. 
                
                $d$ : Dimension of vector representation of the binding site.\\
                These two comparison techniques are used in the sensitivity analysis of this method.

    \subsection{Sensitivity Studies}
    
        \sitevec\ algorithm is evaluated on three types of perturbations of sites for assessing its sensitivity.
        The first evaluation is to perform sanity test with respect to rotation.
        The second study is is to induce perturbations in atomic coordinates and evaluate robustness.
        The third study is to assess the robustness of vectorization with respect to minor changes in residue compositions.
        These three types of perturbations can be expected in any naturally occuring binding sites across protein families and to account for effects of molecular dynamics in protein structures in a living system.
    
    \subsubsection{Sensitivity analysis based on reference frame rotations}
        
            Translation and rotation of coordinates of atoms in a binding sites do not affect the relative positions and topology of atoms.
            Any site representation algorithm should be robust to these types of trivial reference frame changes.
            Any algorithm that directly consumes a binding site as input, faces two challenges.
            The first challenge is the values of coordinates based on the reference frame.
            The second challenge is input order of atoms.
            
            A Deep learning-based method, DeepDrug3D \cite{pu2019deepdrug3d} uses PCA \cite{abdi2010principal} to first determine 3 axes and then re-orient and re-position points about the axes and their center.
            This step is performed as a pre-processing step to {\it normalize} the input.
            
            Methods based on pairwise distances between atoms do not get affected by reference frame changes in the input.
            Fuzcav \cite{weill2010alignment} is one method that enumerates triplets of $C_{\alpha}$ atoms and computes the histogram of number of such triplets resulting in a 4833-dimensional vector.
            This method is robust to reference frame variations.
            The method by Nakumara and co-workers \cite{nakamura2016protein} improves on top of Fuzcav by dimensionality reduction using PCA.
            
\begin{figure*}[h!]
            \begin{multicols}{2}
               \subfloat[]{\includegraphics[scale=0.51]{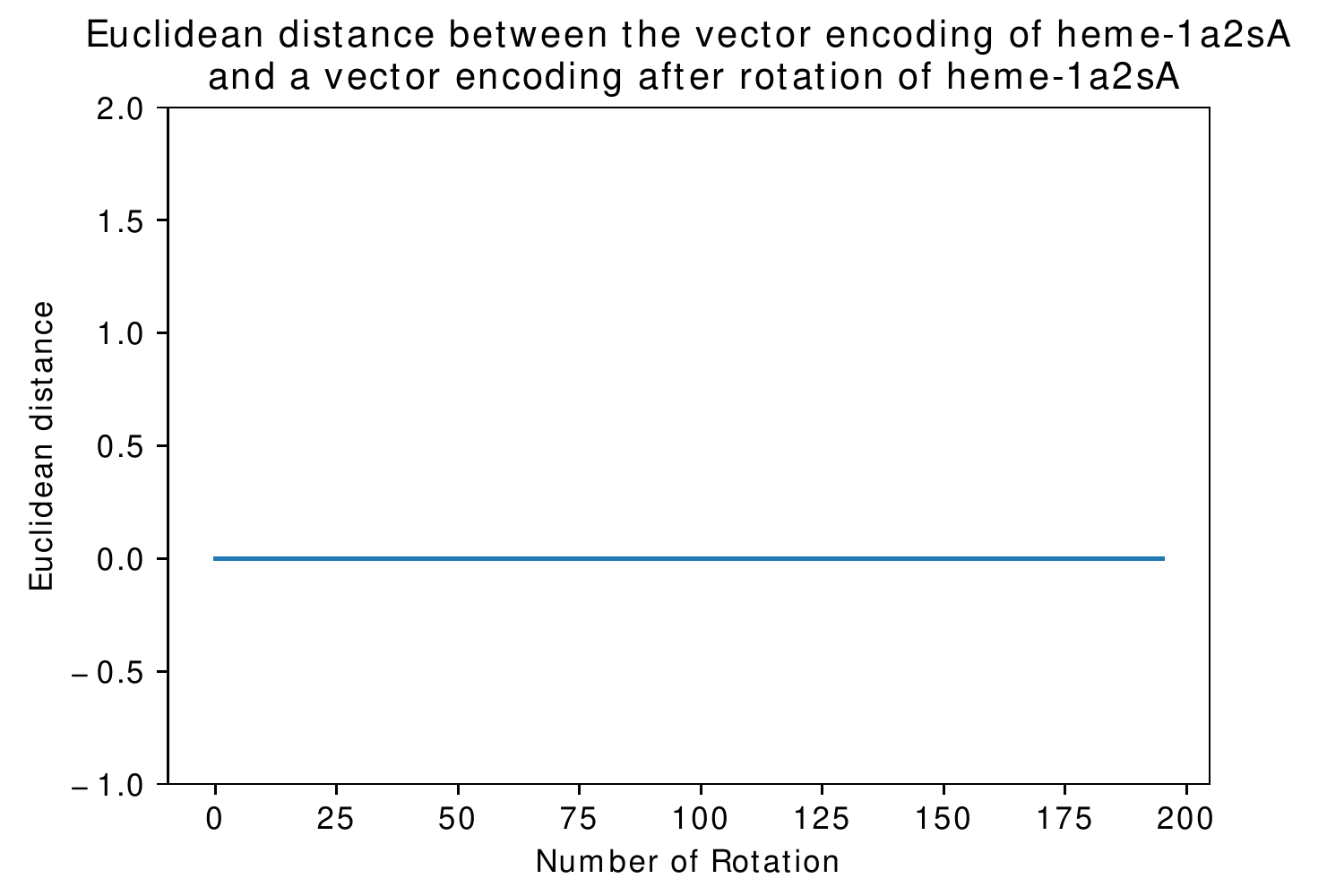}\label{fig:rotation}}
                 \subfloat[]{\includegraphics[scale=0.51]{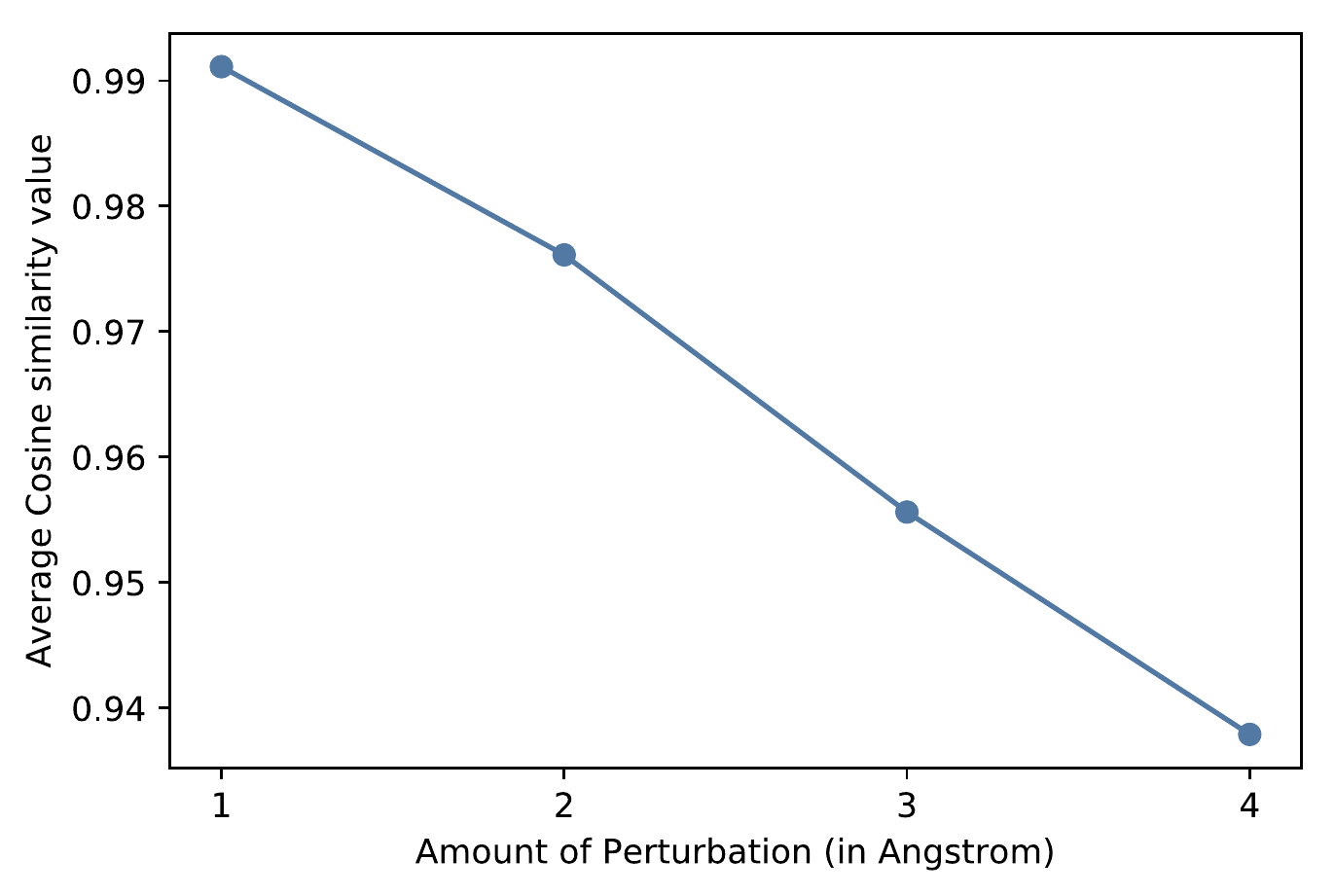}\label{fig:perturbationStudy}}
           \end{multicols}
            \begin{multicols}{2}
                 
                 \subfloat[]{\includegraphics[scale=0.55]{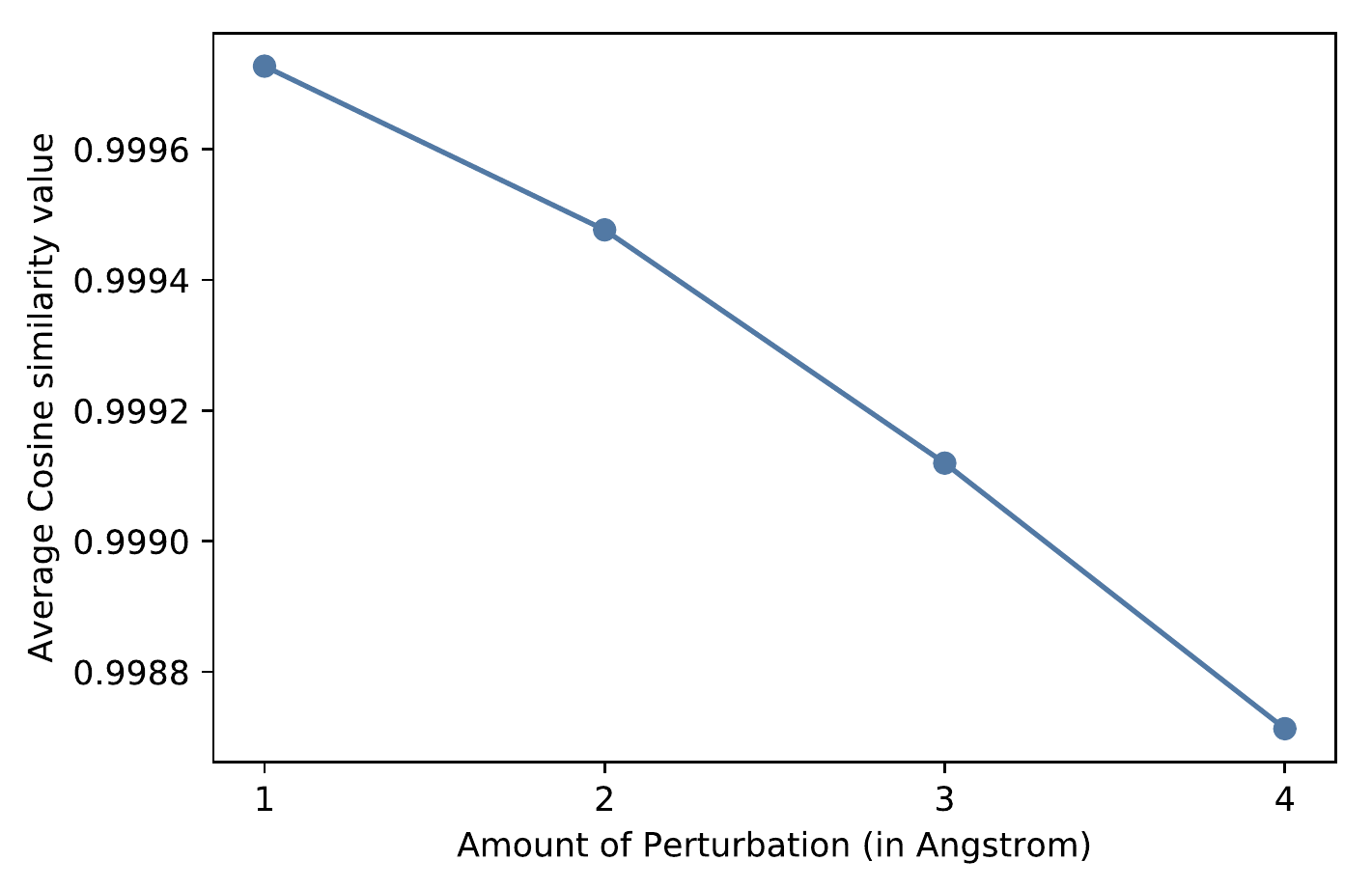}\label{fig:cgammaperturbationStudy}}
                 \subfloat[]{\includegraphics[scale=0.55]{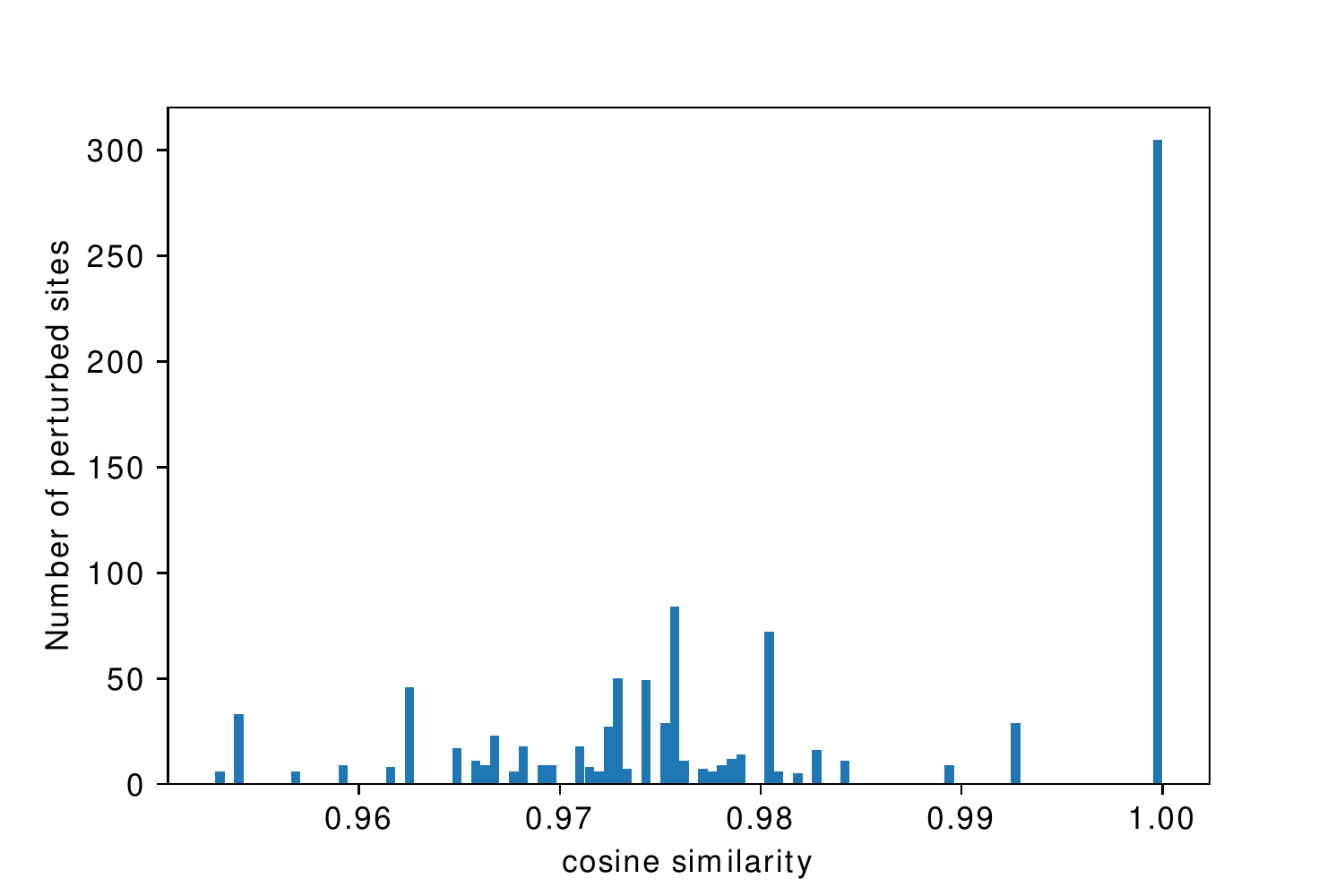}\label{fig:resPerturb}}
            \end{multicols}
            \caption{
            (a) Euclidean distance between vectors of rotated and original heme-1a2sA binding sites. The x-axis is the number of rotation, and the y-axis denotes euclidean distance. (b) Sensitivity analysis based on perturbation of atomic coordinates. The x-axis denotes the amount of perturbation, and the y-axis is the {\bf \it average} cosine similarity value. (c) Sensitivity analysis based on perturbation of centroid of side-chain atoms. The x-axis denotes the amount of perturbation, and the y-axis is the {\bf \it average} cosine similarity value. (d) Sensitivity analysis based on residue type perturbation. This is a histogram where x-axis denotes the cosine similarity values, and the y-axis corresponds to the number of perturbed sites.}
        \end{figure*}

            \sitevec\ algorithm is also a pairwise distance-based method there by inherently robust to reference frame changes in the input.
            We have considered this as a first and sanity test to compute differences between vectors generated for various rotations of a given site.
            As expected, the vectors are identical irrespective of changes in reference frame for a given site.
            We chose heme-1a2sA binding site and generated 250 random rotations of the same.
            Euclidean distances between vectors of the original site before rotation and those of rotated versions are all zero as desired (Figure \ref{fig:rotation}).
            This experiment demonstrates reference frame invariance of \sitevec\ as required.
         
    \subsubsection{Sensitivity analysis based on perturbation of coordinate of atoms of the binding site}
    
        Any vectorization method should be robust to minor perturbations in coordinates of atoms of a binding site.
        In any crystal structure, it is not uncommon to see minor changes in side-chain conformations of the same types of sites across different proteins.
        In this regard, we have carried out a sensitivity analysis of \sitevec\ with respect to perturbations to atomic coordinates of the site.
        
        Natural sites are not expected to be perturbed much due to constraints on bond lengths and atomic forces.
        This poses a challenge in obtaining a natural dataset of sites with atomic perturbations.
        However, synthetically introducing atomic perturbations introduces violations of bond lengths and atomic radii constraints.
        Using molecular dynamics solvers to generate perturbations is also a costly process when whole protein needs to be considered for structural constraints.
        In order to overcome these challenges and introduce a practically feasible way of studying robustness of vectorization to atomic perturbations, we have considered $C_\alpha, C_\beta$ and $\gamma$ coordinates directly (Algorithm~\ref{alg-coord-perturb}).
        Vectors are computed for each of the perturbed versions of a given binding site and are compared against the vector of the unperturbed site.
        It is expected that the cosine similarities are all high between the vectors and only decrease slowly with respect to the amount of perturbation.
        
        We have considered a Heam binding site (1a00-HEM) as an example to assess sensitivity to perturbations from the PLIC dataset.
        As many as 1000 random perturbations are generated using (Algorithm~\ref{alg-coord-perturb}) between 1 \AA \hspace{0.5mm} to 4 \AA \hspace{0.5mm} uniformly in steps of 1 \AA.
        It has been observed that most of the perturbed sites demonstrated very high cosine similarity to the unperturbed sites \ref{fig:perturbationStudy}.
        
        It is also interesting to see how the side chain orientations alone would affect the similarity of vectors between unperturbed and perturbed variations.
        However, molecular dynamics simulation of the entire protein to generate valid side-chain orientations is a costly process and also does not generate many variations.
        We have devised an equivalent experiment to fix $C_\alpha$ and $C_\beta$ and varied only $C_\gamma$ to roughly capture movements of side-chain atoms.
        Given that Site2Vec considers centroid of the side chain atoms, much of the variations in atomic coordinates would be averaged out and the movement of centroid is only expected to be very small.
        We have carried out perturbation studies of $C_\gamma$ coordinates by generating as many as 1000 sites for the Heam binding sites (1a00-HEM).
        The results of the comparison of the unperturbed vector to the perturbed variations resulted in very high cosine similarity values as desired (Figure~\ref{fig:cgammaperturbationStudy}).
        While only $C_\gamma$ points are considered, the $N$ will be $\approx N/3$ in (Algorithm~\ref{alg-coord-perturb}).
        This experiment demonstrates that \sitevec\ is tolerant to positional perturbations of site atoms under natural dynamic conditions of a living cell by capturing similarities between sites even after mild conformational changes.
        
        \begin{algorithm}
        \caption{Algorithm for perturbation of coordinates}
        \label{alg-coord-perturb}
        \begin{algorithmic}[1]
            \STATE Given a net amount of perturbation $\rho$
            \STATE Let $N$ be total number of points in a site (including $C_\alpha, C_\beta$ and $C_\gamma$)
            \STATE $\rho^2 = \frac{\sum_{i=1}^{i=N}{|s_i - s'_i|^2}}{N}$ (where $s_i$ and $s'_i$ are $i^{th}$ coordinates of original and perturbed site respectively)
            \STATE $|s_i - s'_i| = \sqrt{\rho^2/3}$
            \STATE Let $p=s_i$ and $q = s'_i$, then, $q_x \approx p_x \pm \rho/\sqrt{3}$ where ($p_x, q_x$ denotes X coordinates)
            \STATE Similarly for other coordinates as well
        \end{algorithmic}
        \end{algorithm}
    
    \subsubsection{Sensitivity analysis based on perturbation of residue types of the binding site}
    
        This is the third type of perturbation study where the robustness of site vectorization is studied with respect to minor changes in the chemical composition of a binding site.
        To this extent, we have considered single residue type perturbation as a way of inducing chemical perturbation.
        First, we have fixed the coordinates of atoms and randomly picked a residue in a binding site and changed its type to generate a perturbed site.
        From the perturbed site vector descriptor is freshly computed based on \sitevec\ algorithm.
        Cosine similarity is computed between the vectors of the unperturbed site and each of the perturbed sites.
        For each residue, one of the 20 amino acids is chosen as its new type.
        The base work for \sitevec\ algorithm uses \cite{yeturu2008pocketmatch} to fundamentally group residues into 5 groups.
        To arrive at a ballpark number on how many perturbations would result in a distinct chemical type, we can assume each group to represent $\approx 4$ amino acids.
        A random type perturbation falls in one of these 5 groups and in a number of 1000 perturbations, we can expect $\approx 200$ distinct sites.
        We have observed that most of the type perturbed sites possessed high cosine similarity to the original site as desired (Figure~\ref{fig:resPerturb}).

    \subsection{Site2Vec as an online Web Service}

          The \sitevec\ algorithm is provisioned as both a stand-alone version and as a web service at \url{http://services.iittp.ac.in/bioinfo/home}.
          The stand-alone version can be downloaded and locally installed to run in a python environment.
          The web services accepts one or more PDB structures as inputs and computes binding sites for all the groups of HETATM records corresponding to ligands in the file.
          Alternatively the user may also specify a PDB id which is internally downloaded or reused on the server side.
          The service maintains a session for which a user waits and downloads the site vectors.
          The web service provides for search option to retrieve similar binding site for a given site, more details can be seen from the manual in the web portal.
          Backend of web ported is developed by Python3 and numpy, Keras, Tensorflow packages are used and
          Python Flask is used as a web framework in the backend, and the front end uses bootstrap.js libraries to facilitate the interactive interface.

\section{Conclusion:}
In this work, we have presented \sitevec, a machine learning-based, reference frame invariant, ligand-independent method for vector embedding of protein-ligand binding sites.
This method encodes a protein-ligand binding site into $d$- dimensional vector.
The vector representation of the site captures both the chemical and structural properties of binding sites.
The quality of vector representations of protein-ligand binding sites is measured in terms of robustness to physical and chemical perturbations of a site and reference frame changes.
The vector embedding mechanism serves as a locality sensitive hashing function where proximal vectors have similar meaning.
We have demonstrated ability of the vector embeddings to capture site similarity as proximity in their vector dimensional space on the PLIC data set.
The algorithm has been rigorously benchmarked against on 23 methods on 10 different data sets on which \sitevec\ stood the top performer in terms of average AUC quaity scores.
\sitevec\ is a highly configurable algorithm on how it generates vector embeddings and we have evaluated configurations on TOUGH-C1 data set and benchmarked for determining best working default parameters.
The algorithm is evaluated on TOUGH-C1 nucleotide and heme sites, ProSPECCTs and ApocS3 and it performed among the top in termrs of accuracy and AUC scores.
The \sitevec\ algorithm also out performed another deep learning alternative today \cite{pu2019deepdrug3d} in terms of speed by about 3000 times and out performed or matched its performance on reported data sets.
The \sitevec\ algorithm is an unsupervised methodology which does not require a pair of sites as it is an computes encoding for a given site {\it in-silos}.
Such an approach is desirable today for {\it site arithmetic} where proximity of sites in vector space and relative positioning is novel with immense applications in structual bioinformatics and drug discovery.
We report here with one of the very first deep learning based vector embedding methods \sitevec\ in the state of the art backed by thorough benchmarking exercises and quality assessments.

 \vspace*{-10pt}

\section*{Acknowledgement}

The research work is carried out with support from DST India as part of SRG/2019/001648.

\section*{References}
\bibliographystyle{unsrt}  
\bibliography{references2}

\end{document}